# An Approach for Thermal Conductivity Measurements in Thin Films: Combining Localized Surface Topography, Thermal Analysis, and Machine Learning Techniques

Mohsen Dehbashi*, Anna Kaźmierczak-Bałata, Jerzy Bodzenta*

Institute of Physics, Center for Science and Education, Silesian University of Technology, Konarskiego 22B, 44-100, Gliwice, Poland

**Abstract**:

This study presents a comprehensive methodology for determining the thermal conductivity (TC) of materials with high reliability. The methodology addresses issues such as surface topographical variations and substrate interference by combining Scanning Thermal Microscopy (SThM) with machine learning (ML) models and normalization techniques. Micro- and nanostructural variations in thin films exacerbate measurement inconsistencies, reducing repeatability and reliability. These interconnected challenges highlight the need for a novel, flexible, and adaptive methodology that can comprehensively address the complexities of thin film characterization while maintaining accuracy and efficiency. In this approach, sample surface was divided into fine spatial grids for localized thermal and topographical measurements. A substrate-thickness factor ($C$ factor) was introduced to account for thickness and substrate effects on thin film TC, and high-performance Random Forest regression was used to predict TC across a broad range of materials. The models were trained on a dataset of 2,352 measurements that covered a wide range of material properties and then validated with an additional 980 measurements. They achieved high predictive accuracy, with a $R^2$ of 0.97886 during training and 0.96630 during testing. This approach addresses instrumental limitations and integrates experimental techniques with computational modeling, providing a scalable framework for a wide range of material science applications.

## 1. Introduction

### 1.1. Importance and Challenges of Thin Film Thermal Measurements

Thin films are crucial in modern technology because of their thermal, electrical, and optical properties [1, 2, 3]. Depending on application, thin films with specific properties are needed to function properly in various devices and systems, including microelectronics, optoelectronics, and energy storage [4, 5, 6]. This allows for precise customization to meet a variety of industrial and research needs.

Thermal conductivity (TC) is a critical parameter that influences heat dissipation, thermal stability, and energy efficiency in devices. Characterization of the thin films TC is a complex and challenging task that has hindered progress in material science and engineering research.

In general, the TC depends on the material composition, but it is strongly influenced by crystalline structure, structural defects and dopants. A separate problem is the measurement of TC. In the case of submicron thick films, the Scanning Thermal Microscopy (SThM) can be used for TC measurement. However, SThM measurements are strongly influenced by topographical characteristics of sample surface. Surface topography influences thermal transport by impeding or facilitating heat flow [7]. Topographical variations can result in localized thermal resistance ($R_{th}^{s_i}$), which affects heat dissipation and performance.

Scanning Thermal Microscopy (SThM) is a precise technique that offers high spatial resolution and can investigate localized thermal properties at the nanoscale [8, 9]. This technique is a type of Scanning Probe Microscopy, enabling simultaneous measurements of local heat transport properties and surface topography with unparalleled precision. In SThM, a specialized thermal probe scans the sample's surface, detecting intricate temperature-induced resistance changes that reveal localized heat flow and the TC. By investigating how variations in surface topography influence local heat transport, researchers can better predict and optimize thermal performance in thin film materials.

Despite these advantages, traditional SThM based TC measurements have certain limitations. Factors such as tip geometry, contact resistance, and variations in surface topography can introduce measurement uncertainties, complicating data interpretation and limiting the reliability of results [10, 11, 12]. These issues are particularly pronounced at the tip-sample interface, where heat transfer is influenced by material-specific characteristics.

At the tip-sample interface, surface roughness plays a critical role in governing heat transfer. Rough surfaces reduce the effective contact area, leading to increased resistance and introducing variability into thermal measurements [13]. This variability complicates the accurate modeling of thermal contact resistance and hinders the ability of theoretical models to generalize across materials with diverse surface characteristics. Consequently, achieving consistent and accurate TC measurements becomes challenging. Theoretical models often fail to account for the intricate interactions at the nanoscale, further underscoring the importance of experimental validation and the development of advanced computational techniques to address these limitations.

Mechanical models, such as the Rabinovich and Derjaguin-Müller-Toporov (DMT) frameworks, also face significant challenges in capturing the complexities of tip-sample interactions [14]. These models rely on assumptions of uniformity and simplicity, often neglecting irregular or intricate surface geometries. These simplifications work well for ideal setups but cannot adapt to diverse probe-sample interactions, especially with non-standard or uneven surface features. This inability to fully represent real-world configurations highlights the need for improved frameworks that integrate the details of nanoscale interactions and enable more reliable analysis.

The limitations of both thermal and mechanical modeling frameworks emphasize the need for continued advancements in SThM methodologies.

Moreover, one of the most prominent challenges remains the interaction between the thin film and its substrate. Substrate materials, which often exhibit thermal properties vastly different from those of the thin film, can distort measurement results by altering heat flow pathways. This issue becomes particularly pronounced in films with thicknesses below the SThM tip radius (~100 nm), where the TC of the substrate plays a more dominant role in shaping the measured thermal response [15, 16].

Other complicating factors include the need to account for critical parameters, such as the effective heat transfer coefficient during convective cooling and the probe-sample boundary thermal resistance ($R_{th,P}^{s_i}$) [9]. These interrelated challenges highlight the need for a novel, flexible, and adaptive methodology that can comprehensively address the intricacies of thin film characterization while maintaining a balance between precision and efficiency.

Such a methodology would not only overcome the inherent limitations of conventional techniques but also open up new possibilities for exploring and understanding the complex thermal behaviours of thin films across a broad spectrum of materials and applications. By integrating advanced measuring tools and approaches, including detailed analysis of surface topography and thermal signals, researchers can achieve more meaningful and reproducible results, paving the way for innovations in material science and technology.

To address these limitations, this work introduces an innovative and comprehensive approach that leverages advanced techniques to address the intricate limitations posed by surface topography, substrate effects, and sample thickness, while integrating machine learning (ML) for calibration, analysis, and predictive modeling. Traditional methods often fall short in capturing the complexity of thin film TC due to limitations in spatial resolution, substrate interference, and the inability to incorporate surface morphology effectively into predictive frameworks. This study bridges these gaps through a comprehensive methodology that combines high-resolution spatial mapping, simultaneous thermal-topographical measurements, normalization strategies, and advanced computational modeling into a cohesive analytical framework.

The methodology begins by implementing a high-resolution spatial grid mapping strategy, dividing a 2 × 2 μm$^2$ area of the thin film surface into a meticulously structured 16 × 16 grid, with each cell measuring approximately 125 × 125 nm$^2$. This subdivision enables the precise capture of localized thermal property variations as well as micro- and nanostructural influences that are often overlooked in conventional approaches. Such fine spatial resolution is crucial for uncovering micro- and nanoscale heat transfer patterns, where even minor irregularities in surface morphology can lead to significant variations in localized thermal conductivity.

At the core of this methodology is the simultaneous acquisition of thermal and topographical data using a Scanning Thermal Microscopy (SThM) probe. This dual-measurement capability ensures that thermal behavior is contextualized with corresponding surface morphology features, enabling precise correlations between surface irregularities and observed thermal responses. Unlike traditional SThM methods, which mostly tend to focus solely on isolated local thermal measurements, this approach integrates thermal signals across the entire micro- and nanoscale grids. This comprehensive integration captures regional interactions, where heat transfer across adjacent cells influences the overall thermal performance of the sample. By analyzing data in this interconnected manner, the methodology avoids the challenges of localized anomalies distorting thermal behavior, providing a more accurate representation of heat transfer dynamics.

Surface topography is incorporated into the analysis as a critical determinant of thermal behavior, playing a pivotal role in governing the efficiency of thermal energy transfer during Scanning Thermal Microscopy (SThM) measurements. The interaction between the probe and the sample surface is heavily influenced by topographical features, which in turn affect key parameters such as the effective heat transfer coefficient ($h^{s_i}$), probe-sample thermal resistance ($R_{th,P}^{s_i}$), and effective probe-cell contact radius ($r^{s_i}$).

The effective heat transfer coefficient ($h^{s_i}$) represents the rate at which thermal energy is exchanged between the probe tip and the sample surface via convective and conductive heat transfer [9]. This coefficient is sensitive to surface topography, with irregular or non-uniform surfaces causing localized disruptions in airflow and thermal gradients. Smooth, uniform topographies typically exhibit higher heat transfer efficiency, while irregular surfaces with sharp peaks or deep valleys may create nano-regions of stagnant air, reducing the overall heat transfer coefficient and introducing variability into the measured thermal properties.

The probe-sample boundary thermal resistance ($R_{th,P}^{s_i}$) quantifies the resistance to heat flow from the tip into the sample surface. This resistance is significantly influenced by the contact quality between the probe and the sample, which depends on the surface morphology. On highly irregular or rough surfaces, the thermal probe tends to make contact primarily with surface peaks or asperities, leaving valleys uncontacted and filled with low-TC air gaps. These air gaps act as insulating barriers, increasing probe-sample thermal resistance ($R_{th,P}^{s_i}$) and distorting the measured TC. In contrast, smoother surfaces facilitate a more consistent and extensive contact area, minimizing probe-sample thermal resistance ($R_{th,P}^{s_i}$) and ensuring accurate heat transfer measurements.

The effective probe-cell contact radius ($r^{s_i}$) defines the physical area of interaction between the thermal probe and the sample surface. This radius is a key parameter because it determines the geometric scale of heat transfer pathways. Surfaces with larger effective contact radii allow for greater heat flux transfer, as the thermal probe interacts with a broader region of the sample surface. However, on rough or highly skewed surfaces, the effective contact radius is often reduced due to poor physical alignment and uneven contact points. This results in localized probe-sample thermal resistance ($R_{th,P}^{s_i}$) spikes and inconsistent measurements across different surface regions.

At the microscale, a 2 × 2 μm² area covered by 16 × 16 pixels (pixel size was 125 × 125 nm²), referred to as the microgrid, is analyzed. The topographic study meticulously examines Root Mean Squared (RMS) roughness ($R_{rms}^{S}$) and skewness ($R_{sk}^{S}$) of sample surface to quantify the statistical characteristics of the thin film's surface texture and understand their influence on thermal transport properties.

RMS roughness ($R_{rms}^{S}$) provides a measure of the average deviation of sample surface heights from the mean plane, effectively capturing the overall amplitude of surface irregularities [17]. This parameter plays a crucial role in determining the thermal contact resistance between the scanning thermal microscopy (SThM) probe and the sample surface. Surfaces with higher roughness values often exhibit reduced effective thermal contact area, leading to increased resistance and less efficient heat transfer pathways.

Complementing this, skewness ($R_{sk}^s$) characterizes the asymmetry of the sample surface height distribution. A positive skewness indicates a prevalence of surface asperities, whereas a negative skewness signifies a dominance of valleys. These topographical variations influence the probe-sample contact quality [18]. Surfaces with a predominance of peaks reduce the actual physical contact area, thereby increasing thermal resistance ($R_{th}^{s_i}$), while valley-dominated surfaces may trap air gaps, forming localized thermal barriers that affect heat transfer.

At the nanoscale, a 375 × 375 nm² area represented by 3 × 3 pixels, called a nanogrid, is analyzed to delve deeper into the localized thermal and topographical characteristics of sample surfaces by analyzing key parameters such as inclination ($M^{s_i}$), standard deviation of surface heights ($\sigma^{s_i}$), and peak-to-valley variations ($\mu^{s_i}$). These nanoscale parameters offer an in-depth understanding of fine-scale irregularities that microscale analysis alone cannot fully capture, providing critical insights into the mechanisms governing localized heat dissipation and resistance pathways.

Inclination ($M^{s_i}$) represents the gradient or slope of surface features at a highly localized level within each nanogrid cell. This parameter indicates the rate of change in surface height over short distances, offering valuable information about the steepness and angular orientation of nanoscale surface features. Steeper inclinations can significantly affect the interaction between the probe and the sample, altering both the pressure distribution and heat flux pathway. High inclinations may create areas where the thermal probe struggles to maintain uniform contact, leading to variations in localized heat resistance and potential measurement artifacts. In contrast, smoother inclinations facilitate more uniform contact areas, ensuring better heat transfer efficiency and measurement consistency.

The standard deviation of surface heights ($\sigma^{s_i}$, mathematically equivalent to $R_{rms}^s$ but here applied to the nanoscale area) quantifies the degree of variability in surface heights within each nanogrid cell [17]. This parameter serves as a measure of surface topography uniformity on a highly localized scale, capturing even the small height deviations that may escape detection in microscale analysis. A higher standard deviation indicates greater surface irregularity, which can lead to non-uniform probe contact and localized heat transfer inefficiencies. These irregularities may cause thermal hotspots or regions of unexpectedly high resistance, ultimately affecting the overall measurement accuracy of the TC.

Peak-to-valley variations ($\mu^{s_i}$) measure the maximum difference between the highest peak and the lowest valley within each nanogrid cell. This parameter is particularly important in identifying extreme surface features that create pronounced discontinuities in the probe-surface contact profile [14]. Large peak-to-valley variations can result in scenarios where the SThM probe interacts predominantly with surface peaks, leaving valleys largely uncontacted and air-filled, creating localized thermal barriers. Conversely, surfaces with minimal peak-to-valley variations promote more uniform probe contact, leading to consistent heat flux pathways and more accurate measurements of nanoscale thermal properties.

The goal of conducted analysis was to find correlation between topographical parameters listed above and measured thermal signals: the thermal signal ratio ($\varGamma_i$) and phase difference ($\Delta\varphi_i$). The thermal signal ratio ($\varGamma_i$) serves as a comparative metric between the thermal resistance of thermal probe in contact with sample ($R_{th}^{s_i}$) and a reference sample ($R_{th}^{n_i}$) at the nanogrid level. It highlights localized deviations in heat transfer efficiency, revealing how nanoscale surface features modulate the material's ability to conduct thermal energy. Similarly, the phase difference ($\Delta\varphi_i$) captures the temporal offset between thermal signals.

To mitigate the variability introduced by probe alignment errors, substrate-specific effects, and material inhomogeneities, the methodology incorporates ML models for both calibration and predictive analysis. ML algorithms are trained on datasets enriched with thermal and topographical parameters, ensuring that the variability arising from experimental imperfections is accounted for systematically. A preliminary correlation analysis, grounded in Spearman's rank correlation coefficients, is employed to identify and filter out low-impact features, thereby refining the dataset and improving computational efficiency. This strategic feature selection ensures that only the most influential parameters are prioritized in the predictive framework. As a result, the ML models are equipped to deliver reliable predictions of TC across diverse sample systems.

To maintain consistency and ensure reproducibility of results, quartz, a material with well-documented and stable TC properties ($1.5 \text{ W} \cdot \text{m}^{-1} \cdot \text{K}^{-1}$ [20]), is employed as a reference. This normalization step serves to standardize thermal signal measurements. By anchoring all thermal measurements to a stable reference, the methodology eliminates discrepancies that often arise in comparative analyses across different thin film samples. Furthermore, a substrate-thickness factor ($C$ factor) is introduced to address the influence of thin film thickness and substrate effects. This parameter scales TC measurements based on the interaction between substrate TC and the thin film's intrinsic properties, particularly in samples where the thickness falls below the critical 100 nm threshold defined by the SThM tip radius.. This correction ensures that the measurements remain both physically meaningful and analytically consistent, even in thin films where substrate interference is typically most pronounced.

The integrity of the dataset is preserved through careful preprocessing, where trends like sample tilt or drift are removed using low-order polynomial background removal. This preprocessing approach eliminates environmental and instrumental noise while retaining the essential characteristics of the measured signals. High-order corrections, which risk introducing artificial distortions, are deliberately avoided to maintain the natural integrity of the dataset. Furthermore, the dynamic resistance data, with its heightened sensitivity to frequency-dependent thermal variations, is leveraged to capture subtle surface signals that static resistance measurements might otherwise overlook [19].

The methodology establishes a multi-scale ML framework that integrates spatial grid mapping, topographical features, thermal signals, and substrate-thickness effect into a unified computational pipeline. By combining fine-scale spatial mapping, dual-data acquisition, quartz normalization, and advanced ML algorithms, this approach overcomes the limitations of traditional TC characterization. It offers deeper insights into the complex interplay between surface topography and heat transfer mechanisms. The resulting framework is not only versatile and adaptable but also scalable, making it suitable for both fundamental research investigations and industrial-scale thin film applications. Through this innovative approach, the study sets a new benchmark for TC characterization, offering a valuable tool for advancing our understanding of thermal behavior in different material systems.

## 2. Experimental Procedures and Methodology

The details of the samples utilized in this study to validate the proposed methodology are summarized in Table 2.1. Commercial indium tin oxide (ITO) thin films were examined after being annealed in various atmospheres, including vacuum, air, oxygen, nitrogen, carbon dioxide, and a nitrogen-hydrogen mixture, all at a temperature of 400 °C. Zinc oxide (ZnO) thin films were studied using samples grown via Atomic Layer Deposition (ALD) with different cycle counts at two deposition temperatures, 100 °C and 200 °C. Additionally, bulk materials such as glass, glassy carbon, silicon carbide, yttrium aluminum garnet, zinc oxide, and polymethyl methacrylate were incorporated into the study. These thermally isotropic materials were selected to improve model calibration, enhance convergence, and refine the TC predictions for thin films.

Thermal conductivity measurements for the thin films referenced in Table 2.1 were conducted using two separate thermal probes (KNT-SThM-2an thermal probe, Kelvin NanoTechnology, Glasgow, UK). The deliberate use of two distinct probes was a key methodological decision to address inherent variability in SThM measurements. Differences in probe geometry and sensitivity can introduce systematic biases in TC measurements. By incorporating data from multiple probes, the ML model becomes less dependent on the specific characteristics of a single probe, enhancing its generalizability. This approach contrasts with purely theoretical models, which often overlook practical measurement inconsistencies. To further mitigate biases, each probe underwent rigorous calibration and normalization against reference materials. Additionally, TC values in Table 2.1 were derived from averaging measurements taken at multiple locations on each sample, reducing the influence of local topographical variations. This multi-position strategy was consistently applied throughout the study to ensure data reproducibility.

A preliminary correlation analysis revealed a significant negative correlation (–0.707, where –1 indicates perfect negative correlation) between the thermal signals ($\Gamma_i$) obtained in this study and the SThM-measured thermal conductivity values ($\kappa^S$) from Table 2.1. This correlation strengthened (–0.726) when $\Gamma_i$ was compared to the ratio $\kappa^S/\sigma^{Si}$, where $\sigma^{Si}$ represents the standard deviation of surface heights within a nanoscale area for the samples. The improved correlation highlights the importance of accounting for surface topography in thermal measurements. The proximity of these two correlation values (–0.707 vs. –0.726) validates the referenced literature's approach of incorporating topological corrections into TC analysis. These findings were fundamental in shaping the ML framework, where $\kappa^S/\sigma^{Si}$ was adopted as the combined output parameter during model training and validation. This integration ensures that topological variability is inherently addressed.

Another important aspect to note regarding Table 2.1 is the effect of annealing on the thickness of ITO thin films. When the film thickness exceeds the SThM tip radius of 100 nm, its influence on thermal conductivity measurements obtained through SThM becomes negligible [16]. While annealing can alter grain size and crystallinity in polycrystalline ITO, X-ray diffraction (XRD) analysis showed only modest grain growth: the reference sample had a mean grain size of ~56 nm, which increased to ~57 nm after vacuum annealing and ~63 nm after $N_2$ annealing. Such minor changes do not significantly affect film thickness, which remained stable at ~170 nm. Although phonon boundary scattering can reduce TC in highly ordered nanoscale films, the polycrystalline nature of ITO and its thickness (>100 nm) diminish this effect.

Based on this, the approach incorporates advanced instrumentation and ML models, trained on a dataset of 2,352 measurements. These models were validated using an additional 980 measurements, making a total of 3,332 measurements. Key statistical metrics are detailed in Table 2.2. The primary focus of the study is on the TC of the samples ($\kappa^s$), which indirectly serves as the target parameter through the relationship $\chi = \kappa^s/\sigma^{s_i}$. We will delve deeper into this parameter in subsequent sections. Other key parameters analyzed include the thermal signal ratio ($\Gamma_i$), the substrate-thickness factor ($C$ factor), the phase difference ($\Delta\varphi_i$), the RMS roughness ($R^s_{rms}$), the skewness roughness ($R^s_{sk}$), the inclination ($M^{s_i}$), and the peak-to-valley value ($\mu^{s_i}$). Additionally, similar parameters are considered for the reference, quartz, denoted as $R^n_{rms}, R^n_{sk}, M^{n_i}, \mu^{n_i}, \kappa^n/\sigma^{n_i}$ (where $\kappa^n = 1.5\ \text{W} \cdot \text{m}^{-1} \cdot \text{K}^{-1}$ is TC of reference). Each of these parameters contributes to understanding the complex interaction between surface topography and thermal behavior. Their significance and individual contributions will be explored in detail in the following sections.

It is particularly important to note from Table 2.2 that the parameters for the reference exhibit lower values and significantly less variation than the corresponding parameters for the samples. This highlights the critical role of quartz as the chosen reference, ensuring consistent and reliable normalization of thermal signals. This distinction highlights the importance of parameter stability when selecting a reference. Figure 2.1 provides a visual overview of the systems. The left image presents a topography map, while the right image displays a thermal signal map of the same sample. Together, these visuals reveal how structural variations affect thermal transport across micro- and nanoscale dimensions, offering valuable insights into the material's heat transfer performance.

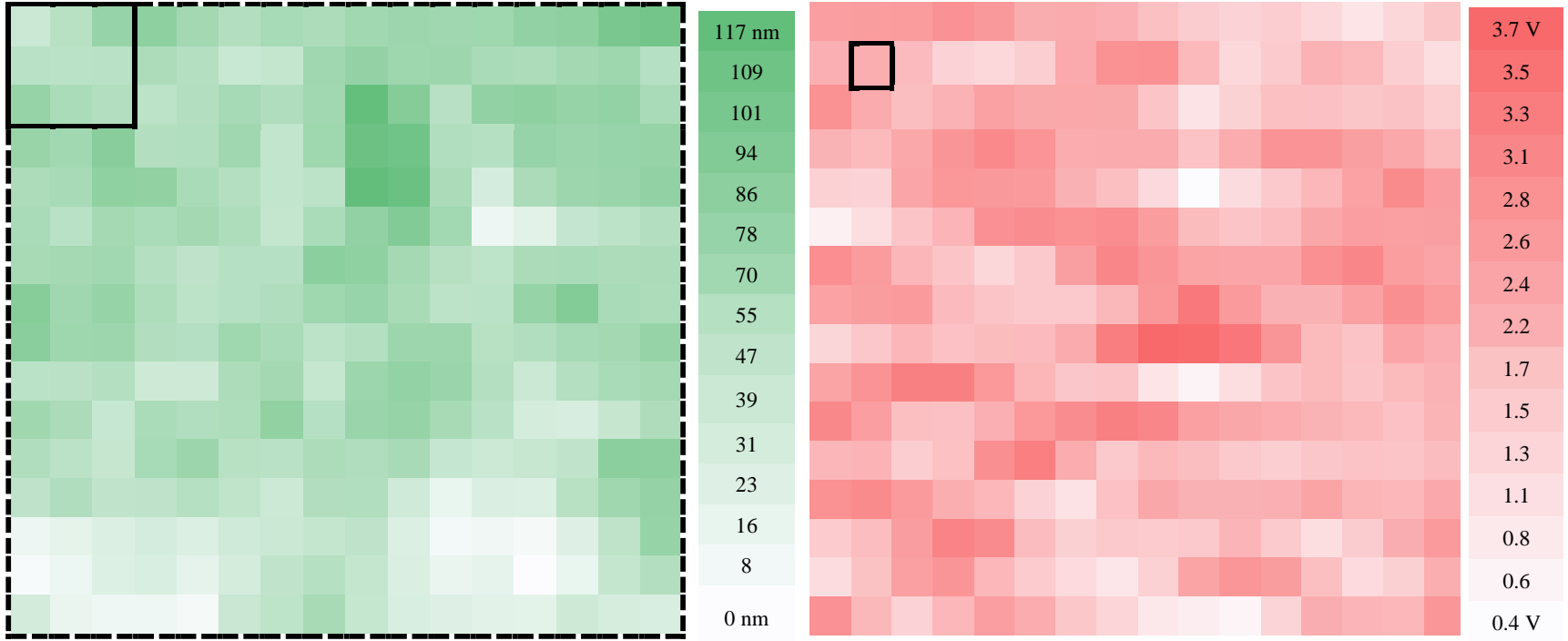

Figure 2.1. The left image displays a topography map of a sample captured on a microscale area (16 × 16 cells). The dotted square highlights the microscale region of 2 × 2 µm² used for surface roughness analysis ($R_{rms}^s, R_{sk}^s, R_{rms}^n, R_{sk}^n$), while the smaller filled square represents a nanoscale subset area of 375 × 375 nm² (3 × 3 cells) for localized topographical characterization ($M^{s_i}, \mu^{s_i}, \sigma^{s_i}, M^{n_i}, \mu^{n_i}, \sigma^{n_i}$). The image on the right shows the thermal signal map of the same material within the same surface region, with the small square indicating the specific cell where thermal signals ($\Gamma_i, \Delta\varphi_i$) were later captured. Note that only the signal for dynamic component of the thermal map is shown here. Two additional components (static component and phase shift map) are required to calculate these parameters. After each measurement, the nanogrid shifts by one cell to the right, and the small square in the thermal map shifts accordingly to capture the next data point, ultimately yielding 196 data points per sample. It should also be noted that these maps represent the "trace" data. Additionally, "retrace" maps are collected, and the data from these two maps are averaged for each cell to reduce noise and enhance the reliability of the dataset.

Table 2.1. Detailed description of the samples used to validate the proposed methodology.

| Sample's number | Sample | Substrate | Layer thicknesses ($nm$) | Actual $\kappa$ (W·m$^{-1}$·K$^{-1}$) |
|---|---|---|---|---|
| 1 | ITO (1)* | Glass | 170 | 6.4 [20] |
| 2 | ITO (2)* | Glass | 170 | 3.5 [21] |
| 3 | ITO (3)* | Glass | 170 | 8.3 [21] |
| 4 | ITO (4)* | Glass | 170 | 10.6 [20] |
| 5 | ITO (5)* | Glass | 170 | 11.8 [21] |
| 6 | ITO (6)* | Glass | 170 | 6.7 [20] |
| 7 | Glass (Bulk) | - | - | 1.1 |
| 8 | Glassy carbon (Bulk) | - | - | 6.3 |
| 9 | Silicon carbide (Bulk) | - | - | 450 |
| 10 | Yttrium aluminum garnet (Bulk) | - | - | 12 |
| 11 | ZnAlO (1)* | Silicon | 110 | 4.29 [22] |
| 12 | Zinc oxide (Bulk) | - | - | 80 |
| 13 | ZnO (2)* | Silicon | 12 | 0.25 [23] |
| 14 | ZnO (3)* | Silicon | 15 | 0.28 [23] |
| 15 | ZnO (4)* | Silicon | 38 | 1.12 [23] |
| 16 | ZnO (5)* | Silicon | 118 | 2.81 [23] |
| 17 | Polymethyl methacrylate (Bulk) | - | - | 0.17 |

* ITO (1) to ITO (6) were annealed in different atmospheres: oxygen, carbon dioxide, vacuum, air, nitrogen, and a nitrogen-hydrogen mixture, respectively. ZnO (1) to ZnO (5) were deposited using ALD with cycle counts of 850, 150, 150, 330, and 900, respectively. The deposition temperatures were 200°C, 100°C, 200°C, 200°C, and 200°C, respectively.

Table 2.2. Overview of the data collected and documented during laboratory experiments.

| Variable* | Observations | Minimum | Maximum | Mean | StDev |
|---|---|---|---|---|---|
| $\Gamma_i$ | 3332 | 0.997 | 1.005 | 1.001 | 0.002 |
| $\Delta\varphi_i$ | 3332 | -693.837 | 826.137 | 6.813 | 148.089 |
| $C$ | 3332 | 0 | 73.333 | 11.52 | 25.219 |
| $R^n_{rms}$(nm) | 3332 | 0.326 | 2.523 | 0.664 | 0.480 |
| $R^n_{sk}$ | 3332 | -1.55 | 0.507 | -0.266 | 0.49 |
| $M^{n_i}$ | 3332 | $3.530 \times 10^{-5}$ | $5.583 \times 10^{-2}$ | $2.786 \times 10^{-3}$ | $3.246 \times 10^{-3}$ |
| $\mu^{n_i}$(nm) | 3332 | 0.258 | 16.7 | 1.395 | 1.324 |
| $\kappa^n/\sigma^{n_i}(GW.m^{-2}.K^{-1})$ | 3332 | 0.243 | 19.468 | 4.484 | 1.997 |
| $R^s_{rms}$(nm) | 3332 | 0.265 | 29.700 | 6.075 | 9.209 |
| $R^s_{sk}$ | 3332 | -0.127 | 3.072 | 0.693 | 0.81 |
| $M^{s_i}$ | 3332 | $5.619 \times 10^{-5}$ | 0.457 | $2.813 \times 10^{-2}$ | $5.516 \times 10^{-2}$ |
| $\mu^{s_i}$(nm) | 3332 | 0.189 | 142.200 | 13.220 | 22.540 |
| $\chi(GW.m^{-2}.K^{-1})$ | 3332 | 0.006 | 1681.611 | 32.650 | 108.729 |

* Variables include thermal signal ratio ($\Gamma_i$), phase difference ($\Delta\varphi_i$), and substrate-thickness factor ($C$). For the reference material (quartz), measurements include RMS roughness ($R^n_{rms}$), skewness ($R^n_{sk}$), inclination ($M^{n_i}$), peak-to-valley variations ($\mu^{n_i}$), and the ratio $\kappa^n/\sigma^{n_i}$, where $\kappa^n$ represents the thermal conductivity of the quartz and $\sigma^{n_i}$ denotes the standard deviation of surface heights for quartz. Corresponding parameters for the sample are also provided: Root Mean Squared (RMS) roughness ($R^s_{rms}$), skewness ($R^s_{sk}$), inclination ($M^{s_i}$), peak-to-valley variations ($\mu^{s_i}$), and the ratio $\chi = \kappa^s/\sigma^{s_i}$ (target variable), where $\kappa^s$ represents the thermal conductivity of the sample and $\sigma^{s_i}$ denotes the standard deviation of surface heights.

## 2.1. Instrumentation and data acquisition setup

The experiments were conducted using an Atomic Force Microscopy (AFM) system (Park Systems XE-70, Suwon, Republic of Korea) equipped for scanning thermal microscopy (SThM). A specialized SThM probe (KNT-SThM-3an thermal probe, Kelvin NanoTechnology, Glasgow, UK) was used, which can simultaneously collect thermal and topographical signals in contact mode. This dual-functionality probe ensures that both datasets are acquired under identical conditions, enabling correlated analysis of surface features and thermal properties. The integration of thermal and topographical data allows for a comprehensive analysis of the relationship between surface morphology and thermal behavior, providing insights into the mechanisms governing heat transfer at small scales. All measurements were performed in a vibration-damped chamber to eliminate external mechanical disturbances, which could otherwise introduce noise or artifacts into the thermal and topographical data. The chamber temperature was maintained at room temperature (approximately 25°C) to prevent thermal distortions caused by heat generated by the SThM system itself. This control is essential for ensuring that the measured thermal signals are solely attributable to the sample properties and not influenced by external thermal fluctuations. A schematic of the probe interaction with the surface is shown in Figure 2.2, along with two high-correlation parameters: inclination and peak-to-valley distance.

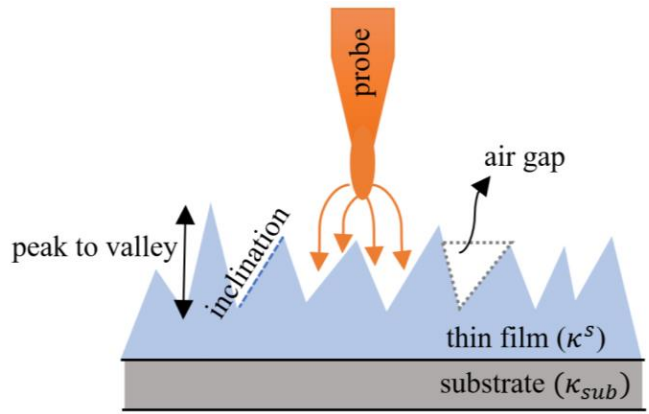

Figure 2.2. Schematic of the SThM probe-surface interaction, highlighting key parameters: inclination and peak-to-valley distance. The probe's dual thermal-topographical sensing enables correlated nanoscale measurements under controlled conditions.

The experiments involved thin film samples and a reference material, quartz. The procedure followed a strict sequence to ensure consistency and reliability. A $2 \times 2$ μm$^2$ area on the quartz was scanned to collect baseline thermal signals, followed by scanning the same area size on the thin film sample to collect its thermal signals. This sequence was repeated, alternating between the quartz reference and different samples. The scan rate was set to 0.05 Hz, chosen to balance measurement resolution and time efficiency, and the setpoint force for the probe was maintained at 1 nN for all samples to ensure uniform contact between the probe and the sample surface, minimizing variability in the data.

To enhance the sensitivity and accuracy of the thermal signal detection, a lock-in amplifier (DSP, Model SR830) was employed. This device isolates the desired signal from background noise by referencing the input signal to a known frequency. This technique is particularly effective in filtering out environmental and electronic interference, ensuring precise extraction of low-amplitude thermal signals. The probe was supplied with both alternating current (AC) at a frequency of 2300 Hz and direct current (DC) using a precision current source (Keithley, Model 6221). The DC current ($I_{DC}$ = 1.8 mA) was used for static resistance measurements under steady-state conditions, providing baseline thermal characterization, while the AC current ($I_{AC}$ = 0.09 mA amplitude) enabled dynamic measurements.

This detailed experimental setup and methodology provide a reliable foundation for the study of thermal transport in thin film samples.

## 2.2. Thermal signals

The effective thermal resistance of each grid cell to heat transfer ($R_{th}^{s_i}$) was compared to the thermal resistance of the reference ($R_{th}^{n_i}$) for the corresponding cell. Quartz was selected as the reference sample due to its stable and well-documented thermal properties. The ratio was expressed as [24]:

$$\frac{R_d^{s_i} - R_s^{s_i}}{R_d^{n_i} - R_s^{n_i}} = \frac{R_{th}^{s_i}}{R_{th}^{n_i}} = \Gamma_i \qquad 2.1$$

Here, $R_d^{s_i} - R_s^{s_i}$ represents the difference in dynamic ($R_d^{s_i}$) and static ($R_s^{s_i}$) electrical resistances of the probe at sample surface cell $i$, while $R_d^{n_i} - R_s^{n_i}$ is the difference in dynamic and static electrical resistances for the reference for the corresponding cell at the i-th cell.

Static resistance was measured under steady-state conditions using a direct current ($I_{DC}$=1.8 mA), while dynamic resistance was measured using an alternating current ($I_{AC}$=0.09 mA – the amplitude) to capture the material's response to time-dependent thermal stimuli. Using a Scanning Thermal Microscopy (SThM) probe, both resistances were measured across all cells. This ratio ($\Gamma_i$) normalized the sample's thermal properties relative to the quartz material, highlighting localized variations in thermal behavior.

The thermal resistance ($R_{th}^{s_i}$) observed by the probe arose from two primary heat exchange mechanisms: (1) convective cooling in the surrounding air and (2) heat flux through the probe-sample contact. This relationship can be expressed as follows [24]:

$$(R_{th}^{s_i})^{-1} = h^{s_i} + \left(R_{th,P}^{s_i} + \frac{1}{4\kappa^s r^{s_i}}\right)^{-1} \qquad 2.2$$

where $h^{s_i}$ is the effective heat transfer coefficient for convective cooling related to cell $i$, $R_{th,P}^{s_i}$ is the probe resistance to heat transfer from the heated volume to the probe–sample contact area, and $r^{s_i}$ is the probe-cell effective contact radius.

## 2.3. Impact of surface topography on thermal parameters

The thermal resistance ($R_{th}^{S_i}$) is highly dependent on the sample's surface topography, as it is governed by key parameters such as $r^{S_i}$, $R_{th,P}^{S_i}$ and $h^{S_i}$ are intrinsically linked to surface topography. Irregular surface topography limited the actual physical contact area between the probe and the sample cell by reducing the effective contact radius ($r^{S_i}$). This occurred as the probe interacted primarily with surface peaks, or asperities, leaving gaps filled with air or other low-TC materials, thereby increasing the probe thermal resistance ($R_{th,P}^{S_i}$). On the other hand, a water meniscus forming in the contact area could decrease this contact thermal resistance. Conversely, smoother surface topography facilitated more uniform contact, enhancing the effective contact area, increasing $r^{S_i}$, and reducing $R_{th,P}^{S_i}$ by minimizing insulating gaps. In addition to $r^{S_i}$ and $R_{th,P}^{S_i}$, the effective heat transfer coefficient for convective cooling ($h^{S_i}$) was also influenced by surface topography. Irregular surface topography can disrupt airflow patterns around the cell, leading to localized variations in $h^{S_i}$. Smoother surface topography promoted more stable heat dissipation through convection, maintaining a consistent $h^{S_i}$ across the measured area. Together, $r^{S_i}$, $R_{th,P}^{S_i}$, and $h^{S_i}$ provided a comprehensive understanding of how surface topography impacts the thermal resistance ($R_{th}^{S_i}$). As derived from the combination of Equations 2.1 and 2.2:

$$(\kappa^s)^{-1} = 4r^{S_i} \left[ \left( \frac{h^{n_i} + \left( R_{th,P}^{n_i} + \frac{1}{4\kappa^n r^{n_i}} \right)^{-1}}{\Gamma_i} - h^{S_i} \right)^{-1} - R_{th,P}^{S_i} \right] \qquad 2.3$$

This equation demonstrates that the TC of the sample $\kappa^s$ is inherently linked to thermal parameters, which, as previously discussed, are significantly affected by topographical characteristics. Importantly, it highlights that not only the intrinsic properties of the sample but also the topographical features of the reference play a critical role in determining accurate TC values. By thoroughly considering both the sample and the normalizing topographical features, this approach significantly enhances the reliability of predictions and provides a deeper understanding of localized thermal behavior across diverse surface conditions.

## 2.4. Microscale and nanoscale perspectives in machine learning prediction

The ML model integrates thermal and topographical parameters derived from both microscale and nanoscale analyses to predict the TC of material systems. The selected features ensure a comprehensive representation of surface characteristics and thermal behavior.

### 2.4.1. Microscale analysis (Microgrid)

At the microscale level, a 2 × 2 μm² area, referred to as a microgrid, is considered for calculating RMS roughness and skewness. This area is divided into a 16 × 16 grid resulting in 256 cells. Each cell captures local surface characteristics and statistical parameters essential for understanding the overall behavior of TC at microscale level.

The RMS roughness ($R_{rms}^s$ and $R_{rms}^n$) represents the roughness amplitude across the sample and reference surfaces, while the skewness of surface roughness ($R_{sk}^s$ and $R_{sk}^n$) indicates whether the surface features of the sample and reference are predominantly peaks or valleys. These parameters provide a statistical overview of surface characteristics on a microscale level. The mathematical formulations for these parameters are as follows:

$$R_{rms}^s = \sqrt{\frac{1}{N}\sum_{i=1}^{N}(r_i^s)^2}, R_{rms}^n = \sqrt{\frac{1}{N}\sum_{i=1}^{N}(r_i^n)^2} \qquad 2.4$$

$$R_{sk}^s = \frac{\frac{1}{N}\sum_{i=1}^{N}(r_i^s)^3}{(R_{rms}^s)^3}, R_{sk}^n = \frac{\frac{1}{N}\sum_{i=1}^{N}(r_i^n)^3}{(R_{rms}^n)^3} \qquad 2.5$$

Here, $r_i^s$ and $r_i^n$ represent the residual at cell $i$, which is the difference between the actual surface height and the predicted height from the fitted regression plane for the sample and reference surfaces, respectively. N=256 denotes the total number of cells in the microgrid.

## 2.4.2. Nanoscale analysis (Nanogrid)

At the nanoscale level, the grid is reduced to a 3 × 3 matrix (9 cells), called a nanogrid, with each nanogrid measuring 375 × 375 nm². This design focuses on more localized and precise thermal and topographical characteristics (Figure 2.1). This approach ensured a precise representation of immediate surface topography, capturing localized variations that could influence thermal resistance ($R_{th}^{s_i}$).

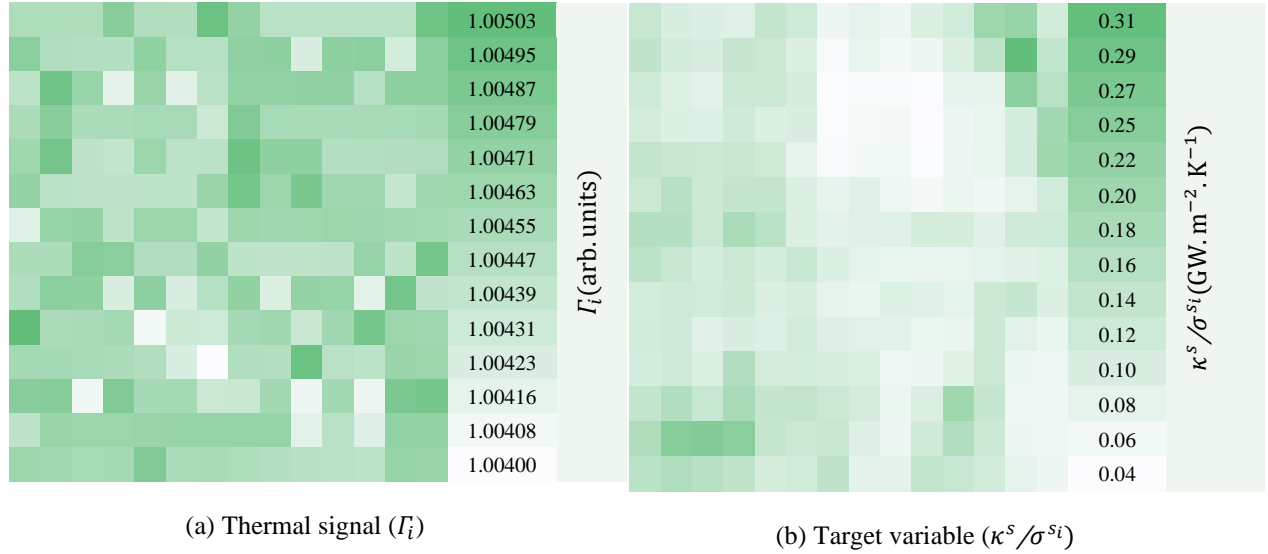

(a) Thermal signal ($\Gamma_i$)  (b) Target variable ($\kappa^s/\sigma^{s_i}$)

Figure 2.3. Spatial distributions of (a) the thermal signal ($\Gamma_i$) and (b) the target variable $\kappa^s/\sigma^{s_i}(GW.m^{-2}.K^{-1})$ on the nanoscale area. The maps correspond to the same region depicted in Figure 2.1.

In the nanoscale analysis, critical parameters including $\Gamma_i$, $(\Delta\varphi_i)$, $M^{s_i}, \mu^{s_i}, \sigma^{s_i}, M^{n_i}, \mu^{n_i},$ and $\sigma^{n_i}$ were captured. The thermal signal ratio ($\Gamma_i$) represents the ratio of thermal signals between corresponding cells on the sample and reference. The phase difference ($\Delta\varphi_i$) captures the temporal phase shift between the thermal signals.

The inclination ($M^{s_i}$) indicates the gradient of surface features within the nanogrid, calculated using surface linear regression coefficients, while the peak-to-valley value ($\mu^{s_i}$) measures the maximum vertical distance between the highest and lowest points the same nanogrid, capturing surface irregularities.

Additionally, the inclination of the reference ($M^{n_i}$) captures the gradient characteristics of the reference surface, while the peak-to-valley value of the reference ($\mu^{n_i}$) measures surface irregularities on the reference grid. The TC-to-standard-deviation ratio ($\kappa^n/\sigma^{n_i}$) describes the relationship between TC and topographical variations in the reference. Here, $\sigma^{n_i}$ represents the standard deviation of surface heights within the nanogrid on the reference surface. It measures how much the heights deviate from the average height ($\bar{Z}$), providing insight into the surface topography across the 9 cells of the grid. Figure 2.3 presents the spatial mappings of the thermal signal ($\Gamma_i$) and the normalized target variable $\kappa^s/\sigma^{s_i}$. The data aligns with the area illustrated in Figure 2.1.

Here border cells have been excluded to mitigate edge-related artifacts, resulting in a refined dataset of 196 points. These maps, along with those in Figure 2.1, show the interplay between thermal characteristics and surface variations, providing insights into their correlated behavior. In Section 2.6, the correlation behavior of these parameters will be quantified through detailed correlation analysis.

## 2.5. A factor for substrate and thickness effects

In this study, we analyzed thin films deposited on various substrates, including silicon and glass, to measure TC using Scanning Thermal Microscopy (SThM). A significant challenge emerged in training the model to distinguish between bulk material samples and thin film samples within the datasets. This distinction was critical because thin film materials exhibit different thermal transport properties compared to their bulk counterparts due to thickness and substrate effects.

To address this, a substrate-thickness factor ($C$ factor) was introduced as follow:

$$C = \begin{cases} \left(\dfrac{b-d}{b}\right)\dfrac{\kappa_{sub}}{\kappa^n}, & d < b \\ 0, & d \geq b \end{cases} \qquad 2.6$$

In this equation, $d$ represents the thickness of the thin film, while $b$ denotes the threshold thickness, typically set at 100 nm (the SThM tip radius). The parameter $\kappa_{sub}$ corresponds to the TC of the substrate.

When the thickness ($d$) is below the threshold ($b$), the factor $C$ scales linearly based on the ratio (($b-d$)/$b$), indicating the extent to which the thin film deviates from the threshold thickness. Additionally, this scaling is weighted by the ratio of the substrate's TC to the reference's TC, $\kappa_{sub}/\kappa^n$.

For thickness values equal to or exceeding the threshold ($d \geq b$), the factor $C$ becomes zero, indicating that the influence of substrate and thickness effects is negligible.

This representation ensures a smooth transition, effectively capturing both geometric and material-specific contributions to the TC behavior of the sample system.

It is important to clarify that the $C$ factor is not a traditional correction factor but is introduced as an input to the ML model. By integrating thickness and substrate effects into a single variable, the model learns boundary conditions and constructs a data-driven representation of $C$ factor during training. This approach eliminates the need for explicit theoretical assumptions about thermal boundary conductance in the model, as the ML framework inherently captures these effects from the dataset.

## 2.6. Spearman's correlation analysis

The correlation coefficients presented in Table 2.3 offer valuable insights into how the target variable ($\chi$) interacts with each input variables. Spearman's rank correlation coefficient ($r_s$, range: -1 to +1) serves as a statistical tool for measuring both the strength and direction of a non-linear association between two variables. Unlike Pearson's correlation, which focuses solely on linear relationships, Spearman's approach relies on the ranks of the data points, making it particularly useful for uncovering linear and non-linear patterns.

The formula for calculating $r_s$ between two ranked variables is given by [25]:

$$r_s = 1 - \frac{6 \sum d_i^2}{N(N^2 - 1)} \qquad 2.7$$

In this equation, $d_i$ denotes the difference between each pair of the ranked variables, while N represents the total number of paired observations. The subsequent analysis interprets the observed Spearman correlation coefficients, explaining the relationships between the target variable ($\chi$) and the various input features. In this table, $\Gamma_i$ displays the strongest negative correlation at -0.726, highlighting its dominant role in influencing $\chi$. Following this, the variables $M^{s_i}$ and $C$ exhibit correlation values of -0.572 and -0.518, respectively, indicating a moderate negative association with $\chi$. Additionally, $R_{rms}^s$ (-0.505) and $\mu^{s_i}$ (-0.609) show notable negative correlations, suggesting a meaningful contribution to the observed variations in $\chi$ On the other hand, variables such as $\Delta\varphi_i$ (-0.022) and $R_{sk}^n$ (-0.005) exhibit weak or negligible correlations, implying minimal influence on $\chi$. In contrast, positive correlations are observed for $R_{sk}^s$ (0.228) and $\kappa^n/\sigma^{n_i}$ (0.098), although their overall impact remains relatively modest. Collectively, these results clarify which variables exert the most significant influence on the target variable. They also emphasize the importance of normalization, as represented by $\Gamma_i$, in amplifying critical relationships while reducing noise from less impactful factors. The correlation factor for $R_d^{s_i} - R_s^{s_i}$ is calculated as -0.024, indicating a weak correlation. However, normalization significantly transforms this relationship, as reflected by the correlation factor of -0.726 for $\Gamma_i$. This substantial change highlights the pronounced effect of normalization in amplifying the underlying relationship and reducing the impact of non-systematic variations.

Furthermore, it is worth noting that all the correlation factors for the sample materials are acceptably higher than those for the reference. This observation aligns with expectations since these parameters are specifically intended for predicting the target variable for the sample material. By ensuring that the correlation factors for the sample materials surpass those of the reference, the analysis confirms the appropriateness of the input features used for prediction. Such differentiation is crucial for accurately modeling the target variable and achieving reliable predictive performance.

Table 2.3. Analysis of the connection between the target variable ($\chi$) and each individual input variable[*].

| | $\Gamma_i$ | $\Delta\varphi_i$ | $C$ | $R^n_{rms}$ (nm) | $R^n_{sk}$ | $M^{n_i}$ | $\mu^{n_i}$ (nm) | $\kappa^n/\sigma^{n_i}$ (GW.m$^{-2}$.K$^{-1}$) | $R^s_{rms}$ (nm) | $R^s_{sk}$ | $M^{s_i}$ | $\mu^{s_i}$ (nm) |
|---|---|---|---|---|---|---|---|---|---|---|---|---|
| $\chi$ (GW.m$^{-2}$.K$^{-1}$) | -0.726 | -0.022 | -0.518 | -0.270 | -0.005 | -0.116 | -0.084 | 0.098 | -0.505 | 0.228 | -0.572 | -0.609 |

[*]The parameters include thermal signal ratio ($\Gamma_i$), phase difference ($\Delta\varphi_i$), and substrate-thickness factor ($C$). For quartz, measurements include RMS roughness ($R^n_{rms}$), skewness ($R^n_{sk}$), inclination ($M^{n_i}$), peak-to-valley variations ($\mu^{n_i}$), and the ratio $\kappa^n/\sigma^{n_i}$, where $\kappa^n$ is thermal conductivity and $\sigma^{n_i}$ is surface height standard deviation. For the sample, parameters are RMS roughness ($R^s_{rms}$), skewness ($R^s_{sk}$), inclination ($M^{s_i}$), peak-to-valley variations ($\mu^{s_i}$), and the target variable $\chi = \kappa^s/\sigma^{s_i}$, where $\kappa^s$ is thermal conductivity and $\sigma^{s_i}$ is surface height standard deviation.

## 3. Machine Learning Model Development

The ML model was developed using various features to predict TC. To address the challenge of limited unique TC values, a continuous multiplicative variable ($\chi$) was introduced as the target output, ensuring the model treated the problem as a continuous regression task. This approach allowed the model to predict TC values across a range, including unseen variables, significantly enhancing its predictive accuracy.

Since TC is inherently a continuous property, misclassifying it as a categorical problem could hinder the model's ability to predict intermediate values between discrete points. The relationship is defined by the model equation:

$$\chi = f_{ML}(\Gamma_i, C, \Delta\varphi_i, R_{rms}^s, R_{sk}^s, M^{s_i}, \mu^{s_i}, R_{rms}^n, R_{sk}^n, M^{n_i}, \mu^{n_i}, \kappa^n/\sigma^{n_i}) \qquad 3.1$$

Where,

$$\chi = \kappa^s/\sigma^{s_i} \qquad 3.2$$

By aligning the output variable with physical measurements from the SThM probe, the model ensured consistency and accuracy in its predictions.

The predictability of the standard deviation of heights ($\sigma^{s_i}$) in this mode as part of output can be indirectly inferred from the thermal signal captured by the SThM probe ($\Gamma_i$), enhancing the model's ability to generalize and validate its predictions. Additionally, the term $\kappa^s r^{s_i}$ in Equation 2.2 serves as a significant target, with $r^{s_i}$ indirectly linked to $\sigma^{s_i}$ as previously discussed, further solidifying the model's physical foundation. Topographical features, such as slope and peak-to-valley variations in the topography map, also have a measurable impact on $\sigma^{s_i}$, contributing to a more comprehensive prediction framework. Furthermore, the reasonable correlation factors between the inputs and the output ($\kappa^s/\sigma^{s_i}$) make it a good choice as a target, as it is essential for validating the model's continuous nature. This supports the use of regression as a more appropriate approach than categorization for capturing the inherently continuous behavior of TC. Finally, by considering both geometrical (e.g., topographical features) and thermal properties in the analysis, the model achieves a more accurate understanding of the system, ensuring meaningful predictions. These arguments support choosing $\chi$ as the output for the ML model. It should be mentioned that for the final prediction of TC using Equation 3.2, $\sigma^{s_i}$ is needed, , which is already measured using information from topography to decouple the TC of the sample from the predicted $\chi$.

The ML model ($f_{ML}$) integrated physical principles, such as thermal signals and surface topography effects, with computational techniques to generalize relationships among the input features. It also should be mentioned that both microscale and nanoscale inputs are combined in the ML model to predict TC. The integration of mentioned parameters ensures that the model captures both large-scale trends and fine-scale variations. This multi-scale approach effectively accounts for surface variations and substrate effects, resulting in more accurate and reliable predictions of TC across diverse sample systems.

To maximize the standardization and utility of these inputs, all data maps, including static and dynamic resistances, topography, and phase changes, were preprocessed using Gwydion software. As part of this approach, a low-order polynomial background removal technique (order one) was employed to maintain measurement accuracy. This method effectively eliminated background without introducing significant alterations to the original dataset, preserving important characteristics for analysis. Higher-order corrections were avoided to prevent distortion of measured properties.

Notably, among the preprocessed inputs, the dynamic resistance data, owing to its frequency-dependent nature, exhibited greater sensitivity to fine surface signals compared to static resistance. This allowed it to effectively capture subtle variations that static measurements might miss [19]. By applying this consistent preprocessing methodology across all inputs, the integrity of the data was preserved while becoming standardized, ensuring uniformity and reliability for subsequent analysis using this approach.

### 3.1. Ensemble Learning Techniques for Regression

The integration of ML techniques into material science has significantly improved the prediction of material properties, reducing reliance on extensive and costly experimental procedures. Among the various ML models, Random Forest regression and Gradient Boosting regression have demonstrated remarkable accuracy in handling nonlinear relationships within complex datasets [26, 27, 28, 29].

Random Forest regression is an ensemble learning method that constructs multiple decision trees using a bootstrapped dataset and averages their outputs to improve predictive accuracy while reducing overfitting [26]. A key advantage of Random Forest regression is its ability to model nonlinear relationships between input variables and material properties. Unlike traditional regression methods, which rely on predefined mathematical formulas, Random Forest dynamically selects the most relevant features, making it highly adaptable to various materials and testing conditions [26, 27].

A critical parameter in Random Forest regression is the minimum number of samples required for node splitting, which influences the model's complexity and performance. Additionally, the method is robust to noise and can process high-dimensional datasets efficiently. Studies have shown that Random Forest outperforms traditional mathematical models in predicting material properties, achieving high predictive accuracy with minimal computational cost [26, 27]. Furthermore, the integration of Bayesian optimization has enhanced Random Forest's predictive performance by fine-tuning hyperparameters, such as the number of estimators (trees) and maximum tree depth.

Gradient Boosting is a sequential ensemble learning technique that builds weak learners—typically decision trees—in an iterative manner, with each subsequent tree correcting the errors of the previous one. Unlike Random Forest, which constructs trees independently, Gradient Boosting focuses on minimizing residual errors through iterative refinement, making it particularly well-suited for high-precision applications in materials science [28, 29].

Gradient Boosting employs an iterative approach where each new model is trained to correct the residual errors of the previous models, following the principles of gradient descent to minimize the loss function. This ensures progressively higher predictive accuracy while maintaining flexibility in handling complex datasets [28, 29].

While both Random Forest and Gradient Boosting are effective ML techniques, their methodologies differ significantly. Random Forest constructs multiple independent decision trees, averaging their outputs for stability, whereas Gradient Boosting builds trees sequentially, refining predictions iteratively. Random Forest is computationally efficient and robust to noise, making it preferable for large datasets. In contrast, Gradient Boosting requires careful tuning to avoid overfitting but often achieves superior precision in highly complex tasks.

### 3.2. Cross-validation strategies and performance assessment metrics

In ML, evaluating the performance of regression models requires a well-defined approach involving both training and testing stages. The dataset is initially divided into two parts: a training set, used to help the model learn patterns and relationships, and a testing set, reserved as an independent benchmark for evaluating the model's predictive ability after training.

To enhance the model's reliability and reduce the risk of overfitting, a widely adopted method is k-fold cross-validation. In this approach, the training data is divided into *k* equally sized subsets, or "folds." During each iteration, *(k-1)* folds are used for training, while the remaining fold serves as a testing set. This process is repeated until every fold has been used once as a validation set. The results from each iteration are then averaged, providing a more comprehensive and unbiased evaluation while reducing the influence of specific data patterns.

Once the optimal hyperparameters have been identified through cross-validation, the model is subjected to final testing using the unseen testing dataset. This stage delivers an objective measure of the model's predictive performance and its ability to handle new, real-world data effectively.

The performance of regression models during both cross-validation and testing is typically quantified using statistical metrics such as R-squared ($R^2$), absolute average relative deviation (AARD%), root mean squared error (RMSE), and mean absolute error (MAE). These metrics offer clear numerical insights into the model's accuracy and reliability. The mathematical expressions for these evaluation metrics are defined as follows:

$$R^2 = 1 - \left(\sum_{m=1}^{M}(y_m^e - y_m^p)^2 / \sum_{m=1}^{M}(y_m^e - y^{ave})^2\right), \text{ where } y^{ave} = (1/M) \times \sum_{m=1}^{M} y_m^e \quad 3.3$$

$$AARD(\%) = (100/M) \times \sum_{m=1}^{M}\left(|y_m^e - y_m^p|/y_m^e\right) \quad 3.4$$

$$RMSE = \left(\sum_{m=1}^{M}(y_m^e - y_m^p)^2/M\right)^{0.5} \quad 3.5$$

$$MAE = (1/M) \times \sum_{m=1}^{M}|y_m^e - y_m^p| \quad 3.6$$

In this equation, $y_m^e$ and $y_m^p$ represent the experimental (observed) and predicted values of the target, respectively. The term $y^{ave}$ denotes the mean value of the experimental target variable. Lastly, M refers to the total number of data points considered during either the cross-validation or testing phases.

## 4. Results and discussion

This section focuses on presenting and analyzing the results obtained from our study. It provides a detailed examination of the key findings, supported by relevant data, visual representations, and statistical analyses.

### 4.1. Hyperparameter optimization and performance evaluation of models

Grid search was employed for hyperparameter tuning to systematically explore predefined parameter spaces for both Gradient Boosting and Random Forest models, ensuring optimal configurations were identified for each. Table 4.1 presents a comprehensive summary of the tested hyperparameters, their respective search ranges, and the total number of models generated throughout the process.

The model evaluation phase utilized cross-validation, a robust statistical technique that partitions the dataset into multiple subsets, training and validating models iteratively to mitigate overfitting and assess generalization performance reliably. In terms of scale, a total of 279 models were systematically trained and evaluated, reflecting the extensive search conducted across the hyperparameter space. The results demonstrated that the Random Forest model outperformed the Gradient Boosting model in predictive accuracy, a conclusion validated by multiple statistical performance metrics. The final optimized hyperparameter sets for Gradient Boosting and Random Forest are detailed in Table 4.2.

Hyperparameters in Random Forest and Gradient Boosting models are essential for optimizing performance, complexity, and generalization. In Random Forests, the number of estimators determines the total number of trees, with more trees generally improving accuracy but also increasing computational cost. The maximum depth parameter controls the depth of each tree, balancing model complexity and the risk of overfitting. Parameters such as minimum samples split, minimum samples leaf, and maximum features help prevent overfitting by ensuring sufficient data in each node and limiting the number of features considered for splitting. The criterion parameter defines the metric used to evaluate the quality of splits. For regression tasks, common criteria include squared error or absolute error. Additionally, advanced variations like Friedman MSE—a specialized version of Mean Squared Error (MSE) designed for gradient boosting—optimize split quality by considering both the mean and variance of the target values. For count data, the Poisson criterion is tailored to leverage the Poisson distribution, assessing the discrepancy between predicted and observed counts to ensure accurate modeling of discrete, non-negative outcomes. Finally, the bootstrap parameter introduces additional randomness by sampling data with replacement, which can improve model performance.

In Gradient Boosting, the number of estimators and maximum depth similarly influence accuracy and complexity. Minimum samples split, minimum samples leaf, and maximum features help avoid overfitting. The learning rate adjusts each tree's contribution, with lower rates requiring more trees for better generalization. The subsample parameter controls the fraction of samples used for training each tree. The loss function defines the objective to be minimized during training. For regression tasks, common choices include squared error for standard regression, Huber loss for robust regression (less sensitive to outliers), quantile loss for predicting specific percentiles, or absolute error for median-focused regression. Additionally, the alpha parameter serves as a regularization term, helping to control the model's complexity and prevent overfitting.

Table 4.1. The hyperparameters of the ML models were explored and analyzed in this study.

| Model's name | Checked hyperparameters | Investigated range | Number of designed models |
|---|---|---|---|
| Random Forest | Number of estimators | 27-4975 | 141 |
| | Maximum depth | 2-998 | |
| | Minimum samples split | 2-198 | |
| | Minimum samples leaf | 1-199 | |
| | Maximum features | 0.11287-0.99951 | |
| | Criterion | Squared error, absolute error, Friedman mse, Poisson | |
| | Bootstrap | True - False | |
| Gradient Boosting | Number of estimators | 22-4960 | 138 |
| | Maximum depth | 2-993 | |
| | Minimum samples split | 7-200 | |
| | Minimum samples leaf | 3-200 | |
| | Maximum features | 0.10153-0.99678 | |
| | Learning rate | 0.01003-0.98148 | |
| | Subsample | 0.12918-0.99875 | |
| | Loss function | Squared error, Huber, quantile, absolute error | |
| | Alpha (Regularization Term) | 0.10124-0.99320 | |

Table 4.2. The optimal hyperparameters selected for each ML model and their associated prediction accuracy.

| Model's name | Tuned hyperparameters | Uncertainty index | Cross-validation training phase | Cross-validation testing phase |
|---|---|---|---|---|
| Random Forest 'a' | Number of estimators = 1986<br>Maximum depth = 923<br>Minimum samples split = 18<br>Minimum samples leaf = 1<br>Maximum features: 0.866<br>Criterion: squared error<br>Bootstrap = False | RMSE | 14.353 | 23.986 |
| | | $R^2$ | 0.98254 | 0.95205 |
| | | AARD% | 6.995 | 14.276 |
| | | MAE | 2.545 | 4.598 |
| Random Forest 'b' | Number of estimators = 2624<br>Maximum depth = 359<br>Minimum samples split = 6<br>Minimum samples leaf = 14<br>Maximum features: 0.961<br>Criterion: absolute error<br>Bootstrap = False | RMSE | 24.445 | 32.770 |
| | | $R^2$ | 0.94747 | 0.92009 |
| | | AARD% | 13.738 | 17.545 |
| | | MAE | 3.720 | 5.519 |
| Random Forest 'c' | Number of estimators = 2306<br>Maximum depth = 419<br>Minimum samples split = 6<br>Minimum samples leaf = 15<br>Maximum features: 0.952<br>Criterion: absolute error<br>Bootstrap = False | RMSE | 25.528 | 30.673 |
| | | $R^2$ | 0.94268 | 0.92791 |
| | | AARD% | 15.477 | 18.304 |
| | | MAE | 4.153 | 5.599 |
| Gradient Boosting | Number of estimators = 2559<br>Maximum depth = 95<br>Minimum samples split = 92<br>Minimum samples leaf = 181<br>Maximum features: 0.923<br>Learning rate = 0.013<br>Subsample = 0.846<br>Loss function: Quantile<br>Alpha = 0.294 | RMSE | 103.205 | 82.509 |
| | | $R^2$ | 0.17953 | 0.23259 |
| | | AARD% | 23.396 | 26.891 |
| | | MAE | 23.107 | 19.444 |

## 4.2. Performance analysis of Gradient Boosting and Random Forest models

As shown in Table 4.2, among all evaluated models, Random Forest 'a' is the most effective, outperforming all assessment metrics in both the cross-validation training and testing phases. The lowest RMSE values among the models are 14.353 for training and 23.986 for testing. These findings demonstrate its remarkable predictive precision and minimal divergence from observed values. The low RMSE is particularly important because it severely penalizes larger errors, increasing the model's reliability for accurate predictions. The model's $R^2$ values (0.98254 in training and 0.95205 in testing) demonstrate its strength, as it explains over 95% of the variance in the target variable during testing. This demonstrates that Random Forest 'a' not only makes precise predictions, but it also generalizes well to new data without overfitting. The average absolute relative deviation (AARD%) values for training and testing are 6.995% and 14.276%, respectively, indicating consistent performance, with predictions closely matching actual values. Furthermore, the Mean Absolute Error (MAE) values of 2.545 in training and 4.598 in testing demonstrate its ability to reduce average prediction errors.

Random Forest 'a' consistently outperforms the other configurations in all metrics. Models 'b' and 'c' have higher RMSE values and lower $R^2$ values, indicating poor predictive accuracy and limited generalization capabilities. These models have elevated AARD% and MAE values, indicating a higher frequency of significant prediction errors and decreased efficacy in identifying data patterns.

When compared to Gradient Boosting, the advantages of Random Forest become more evident. During testing, the Gradient Boosting model had an RMSE of 82.509 and a low $R^2$ value of 0.23259, indicating its inability to accurately model data relationships. Furthermore, the elevated AARD% (26.891) and MAE (19.444) values highlight its lack of reliability and accuracy for this task.

## 4.3. Validation and performance assessment of the optimized Random Forest 'a'

Using the previously optimized hyperparameters, a new Random Forest (model 'a') was trained on a combined dataset, merging both training and testing subsets from the cross-validation phase. Validation was conducted on 980 experimental data points that were not included in the cross-validation process, ensuring the model was tested under realistic, unseen conditions. Predictive accuracy was assessed during the validation (testing) phase, with results detailed in Table 4.3 offering insights into the model's performance metrics. The findings demonstrate the model's generalization ability to predict unseen data effectively.

Table 4.3. Evaluation results of the fine-tuned model 'a' across training and testing phases.

| Metric | Training data | Testing data |
|---|---|---|
| RMSE | 16.033 | 19.222 |
| $R^2$ | 0.97886 | 0.96630 |
| AARD% | 6.319 | 17.487 |
| MAE | 2.511 | 4.448 |

Figure 4.1 illustrates the linear correlation between actual and predicted χ values. Most data points in the training set align closely with the perfect prediction line, demonstrating that the model effectively captured the underlying relationship during training. The testing data points also generally follow the perfect prediction line but exhibit slightly greater variability compared to the training set. For actual χ values exceeding ~ 870 GW.m$^{-2}$.K$^{-1}$, deviations are evident in both the training and testing datasets. This can be attributed to the dataset's higher concentration of samples with low χ values (as detailed in Table 4.4, which shows a greater proportion of samples within lower TC ranges) and a relatively smaller representation of samples with high χ values. Consequently, the observed deviations in this range are reasonable and expected. Notably, the relative error for this range remains relatively small, as illustrated in Figure 4.2. These deviations have minimal impact on practical applications, as TC is determined by averaging numerous data points collected from each sample.

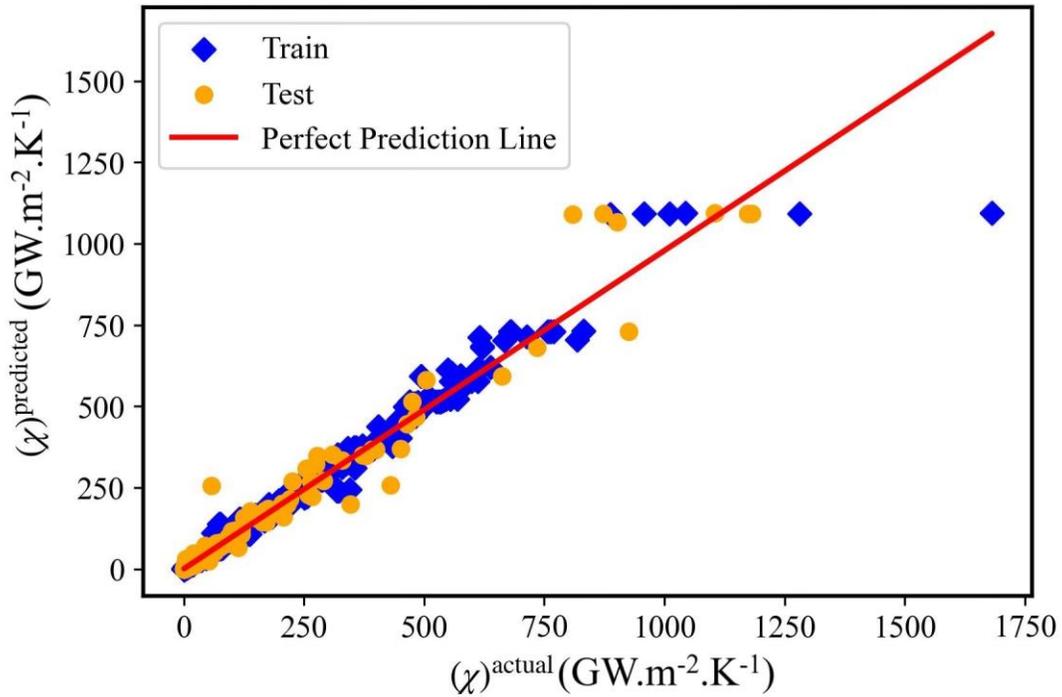

Figure 4.1. Relationship between observed dependent variable values and predicted outputs from the model 'a' for training and testing datasets.

Relative Deviation (RD%) quantifies the signed percentage difference between the predicted and true values, offering insight into the model's performance and bias in predictions. Figure 4.2 illustrates the uncertainty in the predictions of model 'a' in terms of RD% for both training and testing datasets. The majority of RD% values for these datasets are concentrated near the RD = 0 line. For actual χ values greater than ~ 250 $GW.m^{-2}.K^{-1}$, RD% percentages stabilize and remain relatively close to zero. Larger deviations are observed at smaller actual χ values because predictions for low-magnitude values tend to have higher errors. However, these deviations are insignificant in the overall analysis as most predicted data points align closely with the RD = 0 line, indicating more accurate predictions on average. Additionally, the process of averaging TC values smooths out these deviations, ensuring they have minimal influence on the final reported values. This approach enhances the reliability of the results by effectively reducing the weight of outliers or inconsistencies in predictions.

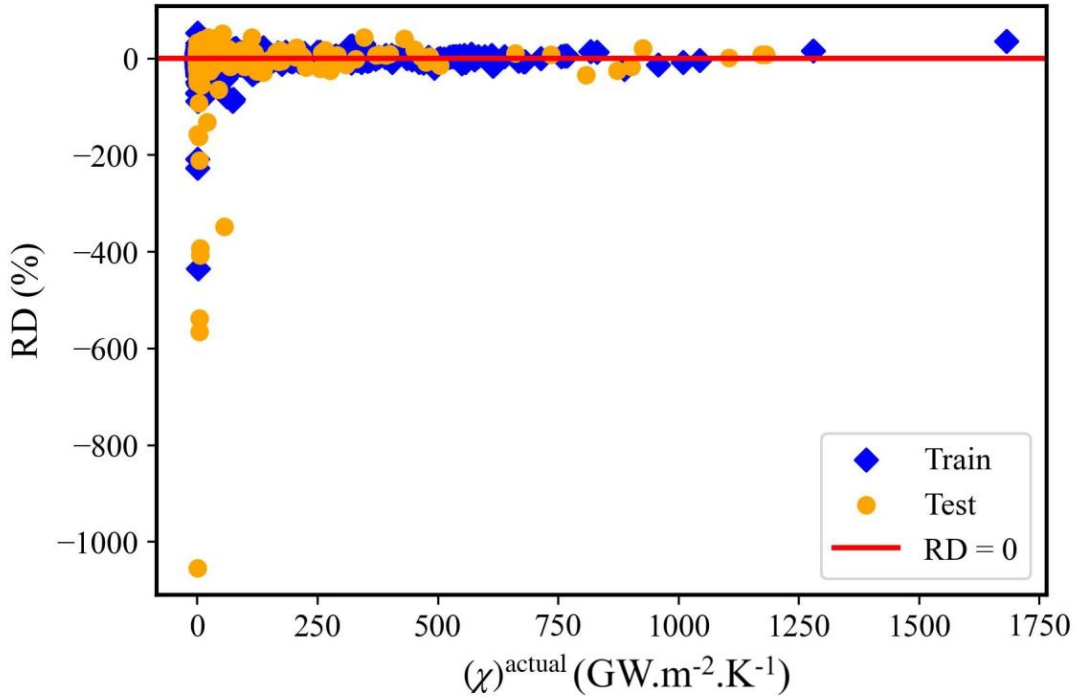

Figure 4.2. The RD% between actual and predicted values of the target variable.

Figure 4.3 illustrates the Gaussian distribution of residuals for both the training and testing datasets, offering insights into the model's predictive efficacy. The training dataset exhibits a mean residual of 0.0, signifying that the model's predictions, on average, correspond precisely with the actual values. The lack of bias highlights the model's capacity to accurately identify the fundamental patterns within the training data. The standard deviation of 16.04 indicates a relatively tight distribution of residuals, reflecting reliable predictions with minimal divergence from the actual values. The testing dataset indicates a mean residual of 0.77, implying a slight positive bias in the model's predictions. This minor bias is acceptable considering the dataset's complexity. The standard deviation of 19.22, while greater than that of the training dataset, highlights the model's ability to adapt to the inherent variability of unseen data.

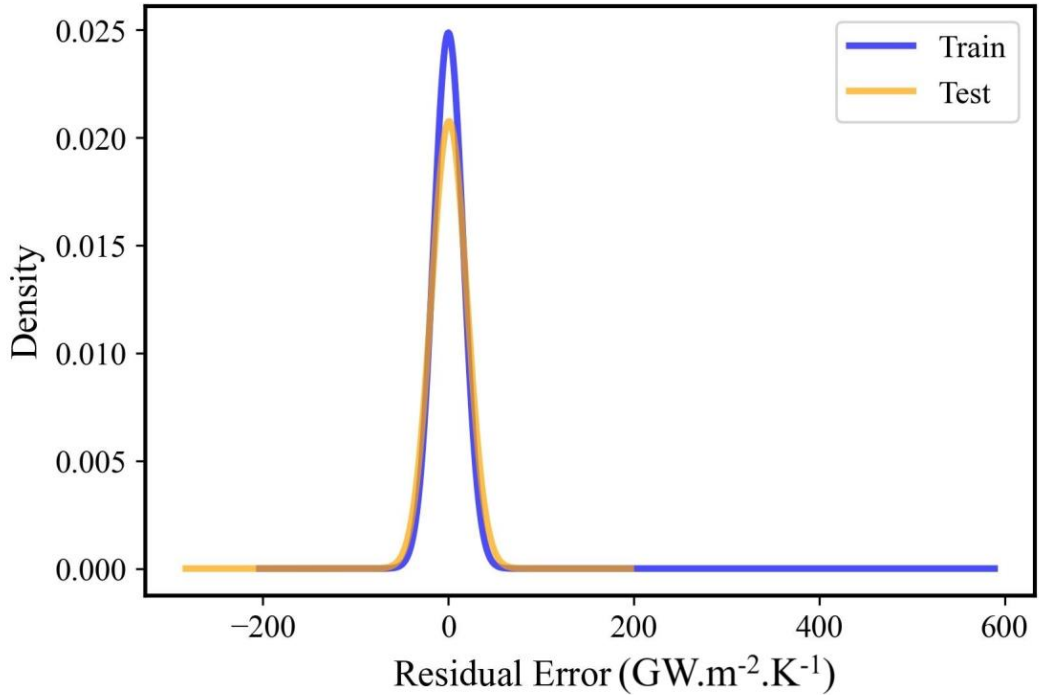

Figure 4.3. Gaussian distribution of residuals for training and testing datasets

The distributions of both datasets provide key insights into the model's performance. The training dataset exhibits a sharp peak and narrow spread, reflecting accurate learning and a strong fit to the data. The testing dataset shows a slightly broader distribution with a marginally lower peak (density of 0.021 compared to 0.025), highlighting the model's ability to handle the complexities of predicting unseen data while maintaining a comparable density profile. The high concentration of residuals near the mean in both datasets demonstrates the model's predictive consistency.

### 4.4. Evaluation of model 'a' predictions for thermal conductivity

Following the prediction of $\chi$ using unseen test data from the previous section, the TC for 17 samples was calculated using Equation 3.2. The predicted TC value for each sample was derived by averaging the multiple predictions generated during the testing phase, ensuring accuracy and reliability in the results. A summary of the predicted and experimental TC values for different samples is provided in Table 4.4. The model demonstrates high predictive accuracy across a wide range of samples, with minimal disparity between actual and predicted $\kappa$ values. For instance, perfect predictions are observed in cases such as Sample 7 (actual: 1.1, predicted: 1.148), Sample 10 (actual: 12, predicted: 12.065), and Sample 15 (actual: 1.12, predicted: 1.122).

Table 4.4. A comparison of the actual and predicted $\kappa$ values for the samples.

| Sample's number | Sample | Actual $\kappa$ (W·m$^{-1}$·K$^{-1}$) | Predicted $\kappa$ (W·m$^{-1}$·K$^{-1}$) |
|---|---|---|---|
| 1 | ITO (1) | 6.4 [20] | 5.303 |
| 2 | ITO (2) | 3.5 [21] | 3.689 |
| 3 | ITO (3) | 8.3 [21] | 8.514 |
| 4 | ITO (4) | 10.6 [20] | 8.452 |
| 5 | ITO (5) | 11.8 [21] | 12.353 |
| 6 | ITO (6) | 6.7 [20] | 10.024 |
| 7 | Glass (Bulk) | 1.1 | 1.148 |
| 8 | Glassy carbon (Bulk) | 6.3 | 6.718 |
| 9 | Silicon carbide (Bulk) | 450 | 463.579 |
| 10 | Yttrium aluminum garnet (Bulk) | 12 | 12.065 |
| 11 | ZnAlO (1) | 4.29 [22] | 4.433 |
| 12 | Zinc oxide (Bulk) | 80 | 87.663 |
| 13 | ZnO (2) | 0.25 [23] | 0.329 |
| 14 | ZnO (3) | 0.28 [23] | 0.285 |
| 15 | ZnO (4) | 1.12 [23] | 1.122 |
| 16 | ZnO (5) | 2.81 [23] | 2.844 |
| 17 | Polymethyl methacrylate (Bulk) | 0.17 | 0.183 |

Furthermore, the model effectively handles an extensive range of $\kappa$ values, from very low (e.g., Sample 17: actual 0.17, predicted 0.183) to extremely high (e.g., Sample 9: actual 450, predicted 463.579), demonstrating its adaptability in capturing trends even for challenging data points. Most samples show a strong correlation between actual and predicted values, proving the model is reliable and works well for different materials and TC ranges. While differences exist (e.g., Sample 12: actual 80, predicted 87.663), they remain within acceptable bounds and can be attributed to challenges in addressing imbalanced cases in the training dataset. As mentioned, the training dataset primarily focuses on relatively low thermal conductivities. In general, the results confirm the approach's reliability and effectiveness in estimating thermal conductivity across a wide array of thermal conductivity ranges.

## 5. Conclusions

This study presents a comprehensive and novel methodology for precisely characterizing the thermal conductivity of thin films, tackling significant issues such as substrate interference, surface topographical variations, and nanoscale irregularities. This method utilizes high-resolution Scanning Thermal Microscopy (SThM) measurements and incorporates ML models with sophisticated normalization techniques to attain remarkable reliability in forecasting thermal conductivity for a wide array of materials. The methodology's ability to generalize effectively is proven by a comprehensive dataset comprising 3,332 measurements, including 2,352 for training and 980 for the final validation stage. The optimized Random Forest regression model demonstrated exceptional predictive performance, achieving a R² of 0.96630 in testing and a minimal mean residual of 0.77, signifying reliable predictions. The residual analysis further illustrates the model's capacity to prevent overfitting. This methodology integrates experimental and computational approaches, creating an adaptable framework for characterizing thin films, thereby holding considerable promise for the advancement of materials science and engineering. Subsequent research can utilize this foundation to investigate further thin-film properties and extend the methodology's application to other fields, fostering additional innovation in next-generation technologies.

**Author Contributions:** M.D.: Machine learning model development and data analysis, Methodology, Data curation and preprocessing, Writing original draft, Investigation. A.K.: Experimental design and execution, Writing and review editing, Project administration. J.B.: Supervision, Instrumentation setup and calibration, Writing and review editing, Final approval. All authors have read and agreed to the published version of the manuscript.

**Acknowledgments:**
The authors acknowledge the ESPEFUM Laboratory at the Institute of Physics – CSE, Silesian University of Technology, for providing access to research facilities. A.K.-B. acknowledges the support from the Silesian University of Technology through the pro-quality grants (Project Nos. 14/030/SDU/10-04-01 and 14/030/SDU/10-27-01).

# References

[1] Zhao, D., Qian, X., Gu, X., Jajja, S. A., & Yang, R. (2016). Measurement techniques for thermal conductivity and interfacial thermal conductance of bulk and thin film materials. *Journal of Electronic Packaging*, *138*(4). https://doi.org/10.1115/1.4034605

[2] Sivaperuman, K., Thomas, A., Thangavel, R., Thirumalaisamy, L., Palanivel, S., Pitchaimuthu, S., Ahsan, N., & Okada, Y. (2023). Binary and ternary metal oxide semiconductor thin films for effective gas sensing applications: A comprehensive review and future prospects. *Progress in Materials Science*, *142*, 101222. https://doi.org/10.1016/j.pmatsci.2023.101222

[3] Adewinbi, S. A., Busari, R. A., Adewumi, O. E., & Taleatu, B. A. (2021). Effective photoabsorption of two-way spin-coated metal oxides interfacial layers: Surface microstructural and optical studies. *Surfaces and Interfaces*, *23*, 101029. https://doi.org/10.1016/j.surfin.2021.101029

[4] Schmitz, J. (2017). Low temperature thin films for next-generation microelectronics (invited). *Surface and Coatings Technology*, *343*, 83–88. https://doi.org/10.1016/j.surfcoat.2017.11.013

[5] Chandekar, K. V., Alkallas, F., Trabelsi, A. B. G., Shkir, M., Hakami, J., Khan, A., Ali, H. E., Awwad, N. S., & AlFaify, S. (2022). Improved linear and nonlinear optical properties of PbS thin films synthesized by spray pyrolysis technique for optoelectronics: An effect of Gd3+ doping concentrations. *Physica B Condensed Matter*, *641*, 414099. https://doi.org/10.1016/j.physb.2022.414099

[6] Amate, R. U., Morankar, P. J., Teli, A. M., Beknalkar, S. A., Chavan, G. T., Ahir, N. A., Dalavi, D. S., & Jeon, C. (2024). Versatile electrochromic energy storage smart window utilizing surfactant-assisted niobium oxide thin films. *Chemical Engineering Journal*, *484*, 149556. https://doi.org/10.1016/j.cej.2024.149556

[7] Dong, L., & Li, Y. (2022). Experimental identification of topography-based artifact phenomenon for micro-/nanoscale thermal characterization of polymeric materials in scanning thermal microscopy. *AIP Advances*, *12*(4). https://doi.org/10.1063/5.0088360

[8] Gucmann, F., Pomeroy, J. W., & Kuball, M. (2021). Scanning thermal microscopy for accurate nanoscale device thermography. *Nano Today*, *39*, 101206. https://doi.org/10.1016/j.nantod.2021.101206

[9] Zhang, Y., Zhu, W., Hui, F., Lanza, M., Borca-Tasciuc, T., & Rojo, M. M. (2019b). A review on Principles and Applications of Scanning Thermal Microscopy (STHM). *Advanced Functional Materials*, *30*(18). https://doi.org/10.1002/adfm.201900892

[10] Martinek, J., Klapetek, P., & Campbell, A. C. (2015). Methods for topography artifacts compensation in scanning thermal microscopy. *Ultramicroscopy*, *155*, 55–61. https://doi.org/10.1016/j.ultramic.2015.04.011

[11] Guen, E., Klapetek, P., Puttock, R., Hay, B., Allard, A., Maxwell, T., Chapuis, P., Renahy, D., Davee, G., Valtr, M., Martinek, J., Kazakova, O., & Gomès, S. (2020). SThM-based local thermomechanical analysis: Measurement intercomparison and uncertainty analysis. *International Journal of Thermal Sciences*, *156*, 106502. https://doi.org/10.1016/j.ijthermalsci.2020.106502

[12] Klapetek, P., Martinek, J., Grolich, P., Valtr, M., & Kaur, N. J. (2017). Graphics cards based topography artefacts simulations in Scanning Thermal Microscopy. *International Journal of Heat and Mass Transfer*, *108*, 841–850. https://doi.org/10.1016/j.ijheatmasstransfer.2016.12.036

[13] Zhang, Q., Zhu, W., Zhou, J., & Deng, Y. (2023). Realizing the Accurate Measurements of Thermal Conductivity over a Wide Range by Scanning Thermal Microscopy Combined with Quantitative Prediction of Thermal Contact Resistance. *Small*, *19*(32). https://doi.org/10.1002/smll.202300968

[14] Guen, E., Chapuis, P., Kaur, N. J., Klapetek, P., & Gomés, S. (2021b). Impact of roughness on heat conduction involving nanocontacts. *Applied Physics Letters*, *119*(16). https://doi.org/10.1063/5.0064244

[15] Guen, E., Chapuis, P., Rajkumar, R., Dobson, P. S., Mills, G., Weaver, J. M. R., & Gomés, S. (2020). Scanning thermal microscopy on samples of varying effective thermal conductivities and identical flat surfaces. *Journal of Applied Physics*, *128*(23). https://doi.org/10.1063/5.0020276

[16] Juszczyk, J., Kaźmierczak-Bałata, A., Firek, P., & Bodzenta, J. (2017). Measuring thermal conductivity of thin films by Scanning Thermal Microscopy combined with thermal spreading resistance analysis. *Ultramicroscopy*, *175*, 81–86. https://doi.org/10.1016/j.ultramic.2017.01.012

[17] McBride, J., & Liu, H. (2020). The Relationship between Contact Resistance and Roughness (Sq) of a Bi-layered Surface using a Finite Element Model. *2020 IEEE 66th Holm Conference on Electrical Contacts and Intensive Course (HLM)*, 176–181. https://doi.org/10.1109/hlm49214.2020.9307833

[18] Trefon-Radziejewska, D., Juszczyk, J., Krzywiecki, M., Hamaoui, G., Horny, N., Antoniow, J., & Chirtoc, M. (2021). Thermal characterization of morphologically diverse copper phthalocyanine thin layers by scanning thermal microscopy. *Ultramicroscopy*, *233*, 113435. https://doi.org/10.1016/j.ultramic.2021.113435

[19] Bodzenta, J., Juszczyk, J., & Chirtoc, M. (2013b). Quantitative scanning thermal microscopy based on determination of thermal probe dynamic resistance. *Review of Scientific Instruments*, *84*(9). https://doi.org/10.1063/1.4819738

[20] Kaźmierczak-Bałata, A., Bodzenta, J., Dehbashi, M., Mayandi, J., & Venkatachalapathy, V. (2022). Influence of post processing on thermal conductivity of ITO thin films. *Materials*, *16*(1), 362. https://doi.org/10.3390/ma16010362

[21] Kaźmierczak-Bałata, A., Bodzenta, J., Szperlich, P., Jesionek, M., Michalewicz, A., Domanowska, A., Mayandi, J., Venkatachalapathy, V., & Kuznetsov, A. (2024). Impact of Annealing in Various Atmospheres on Characteristics of Tin-Doped Indium Oxide Layers towards Thermoelectric Applications. *Materials*, *17*(18), 4606. https://doi.org/10.3390/ma17184606

[22] Kaźmierczak-Bałata, A., Bodzenta, J., & Guziewicz, M. (2019). Microscopic investigations of morphology and thermal properties of ZnO thin films grown by atomic layer deposition method. *Ultramicroscopy*, *210*, 112923. https://doi.org/10.1016/j.ultramic.2019.112923

[23] Kaźmierczak-Bałata, A., Grządziel, L., Guziewicz, M., Venkatachalapathy, V., Kuznetsov, A., & Krzywiecki, M. (2021). Correlations of thermal properties with grain structure, morphology, and defect balance in nanoscale polycrystalline ZnO films. *Applied Surface Science*, *546*, 149095. https://doi.org/10.1016/j.apsusc.2021.149095

[ 24 ] Juszczyk, J., Wojtol, M., & Bodzenta, J. (2013). DC experiments in quantitative scanning thermal microscopy. *International Journal of Thermophysics*, *34*(4), 620–628. https://doi.org/10.1007/s10765-013-1449-4

[25] Xiao, C., Ye, J., Esteves, R. M., & Rong, C. (2015). Using Spearman's correlation coefficients for exploratory data analysis on big dataset. *Concurrency and Computation Practice and Experience*, *28*(14), 3866–3878. https://doi.org/10.1002/cpe.3745

[26] Shingala, B., Panchal, P., Thakor, S., Jain, P., Joshi, A., Vaja, C. R., Siddharth, R. K., & Rana, V. A. (2024). Random Forest Regression Analysis for Estimating Dielectric Properties in Epoxy Composites Doped with Hybrid Nano Fillers. *Journal of Macromolecular Science Part B*, *63*(12), 1297–1311. https://doi.org/10.1080/00222348.2024.2322189

[27] Guo, J., Zan, X., Wang, L., Lei, L., Ou, C., & Bai, S. (2023). A random forest regression with Bayesian optimization-based method for fatigue strength prediction of ferrous alloys. *Engineering Fracture Mechanics*, *293*, 109714. https://doi.org/10.1016/j.engfracmech.2023.109714

[28] Wekalao, J., Srinivasan, G. P., Patel, S. K., & Al-Zahrani, F. A. (2024). Optimization of graphene-based biosensor design for haemoglobin detection using the gradient boosting algorithm for behaviour prediction. *Measurement*, *239*, 115452. https://doi.org/10.1016/j.measurement.2024.115452

[29] K, P. K., Alruqi, M., Hanafi, H., Sharma, P., & Wanatasanappan, V. V. (2023). Effect of particle size on second law of thermodynamics analysis of Al2O3 nanofluid: Application of XGBoost and gradient boosting regression for prognostic analysis. *International Journal of Thermal Sciences*, *197*, 108825. https://doi.org/10.1016/j.ijthermalsci.2023.108825